\documentclass{iau} 
\usepackage{graphicx}

\title[Vohl \etal\ 2016.~~Collaborative visual analytics of radio surveys in the Big Data era] %% give here short title %%
{Collaborative visual analytics of radio surveys in the Big Data era}

\author[D. Vohl, C. J. Fluke, A. H. Hassan, D. G. Barnes \& V. A. Kilborn]   %% give here short author list %%
{Dany Vohl,$^1$
Christopher J. Fluke,$^{1,2}$
Amr H. Hassan,$^1$ 
David G. Barnes$^{2,3}$ \& 
Virginia A. Kilborn$^1$}

\affiliation{$^{1}$Centre for Astrophysics \& Supercomputing, Swinburne University of Technology, 1 Alfred Street, Hawthorn 3122, Australia \\ emails: {\tt dvohl@swin.edu.au, cfluke@swin.edu.au, ahassan@swin.edu.au, vkilborn@swin.edu.au} \\[\affilskip]
$^2$Monash e-Research Centre, Monash University, 14 Alliance Lane, Clayton 3168, Australia \\email: {\tt david.g.barnes@monash.edu} \\[\affilskip]
$^3$Faculty of Information Technology, Monash University, Clayton, Victoria, Australia}

\pubyear{2017}
\volume{325}  %% insert here IAU Symposium No.
\jname{Astroinformatics 2016}
\editors{Massimo Brescia, eds.}
\begin{document}

\maketitle

\begin{abstract}
Radio survey datasets comprise an increasing number of individual observations stored as sets of multidimensional data. In large survey projects, astronomers commonly face limitations regarding: 1) interactive visual analytics of sufficiently large subsets of data; 2) synchronous and asynchronous collaboration; and 3) documentation of the discovery workflow. To support collaborative data inquiry, we present {\tt encube}, a large-scale comparative visual analytics framework. {\tt encube} can utilise advanced visualization environments such as the CAVE2 (a hybrid 2D and 3D virtual reality environment powered with a 100 Tflop/s GPU-based supercomputer and 84 million pixels) for collaborative analysis of large subsets of data from radio surveys. It can also run on standard desktops, providing a capable visual analytics experience across the display ecology. {\tt encube} is composed of four primary units enabling compute-intensive processing, advanced visualisation, dynamic interaction, parallel data query, along with data management. Its modularity will make it simple to incorporate astronomical analysis packages and Virtual Observatory capabilities developed within our community. We discuss how {\tt encube} builds a bridge between high-end display systems (such as CAVE2) and the classical desktop, preserving all traces of the work completed on either platform -- allowing the research process to continue wherever you are.
\end{abstract}

\firstsection % if your document starts with a section,
\section{Common limitations in large-scale radio survey}
%The current count of resolved images of the neutral atomic hydrogen (HI) content in galaxies is in the hundreds. 
Most of the knowledge obtained about neutral atomic hydrogen (HI) in galaxies comes from unresolved observations with large radio telescopes such as the Arecibo, Parkes, and Jodrell Bank radio telescopes. The current data archive contains a few hundreds of resolved HI images. With upcoming next-generation radio telescope facilities such as the Square Kilometre Array (SKA; e.g. \cite{Quinn2015}) and its related pathfinders, e.g. Australian SKA Pathfinder (\cite{Johnston2008ExA}) and the APERTIF upgrade on the Westerbork telescope (\cite{Verheijen2009pra..confE..10V}), thousands of resolved images of the HI distribution of galaxies are expected to be observed. Hence, a revolution in galaxy evolution studies is about to happen, and novel solutions will be needed to be able to make the comparisons required to understand what role hydrogen gas plays in galaxy evolution. 

Hyperspectral images or {\em spectral cubes} are used to study the properties of HI in galaxies. A spectral cube is composed of two spatial dimensions along with a spectral or a velocity dimension. We use the expression `spectral cube {\em survey}' for scenarios where multiple spectral cubes are collected. A primary goal of any spectral cube survey is to identify and investigate similarities and differences between individual sources. While the growth of spectral cubes -- both in number and in size -- within surveys will allow novel science to be undertaken, significant challenges are arising regarding knowledge discovery and analysis. 

With the growth in number of individual spectral cubes within a given survey, new limitations emerge with regard to: 1) interactive visualisation and analysis of sufficiently large subsets of data; 2) synchronous and asynchronous collaboration; and 3) documentation of the discovery workflow. To overcome these challenges, astronomers can borrow techniques from the field of visual analytics -- defined as the science of analytical reasoning facilitated by interactive visual interfaces (\cite{Thomas10.1109/MCG.2006.5}). Visual analytics includes techniques like information synthesis and insight derivation from massive, dynamic, ambiguous, and conflicting data. 

% SECTION 2
\section{{\tt encube}: interactive visual analytics of spectral cubes}
To provide support for collaborative data inquiry in large radio surveys, we introduce \texttt{encube} (\cite{Vohl2015ASPC, 2016arXiv161000760V}), a large scale comparative visual analytics framework. \texttt{encube} is tailored for use with large tiled-displays and advanced immersive environments like the CAVE2 at Monash University. With its 80 stereo-capable screens arranged in a 20 columns $\times$ 4 rows grid configuration, and powered with a $\sim$100 Tflop/s Graphics Processing Unit (GPU)-based supercomputer, the Monash CAVE2 represents a modern hybrid 2D and 3D virtual reality environment. The \texttt{encube} framework is designed to harness the power of high-end visualisation environments for collaborative analysis of large subsets of data from radio surveys. Alternatively, it can also work on standard desktops, providing a seamless visual analytics experience regardless of the number of displays or their arrangement. 

\begin{figure*}[!htb]
\centering
\includegraphics[width=13.1cm]{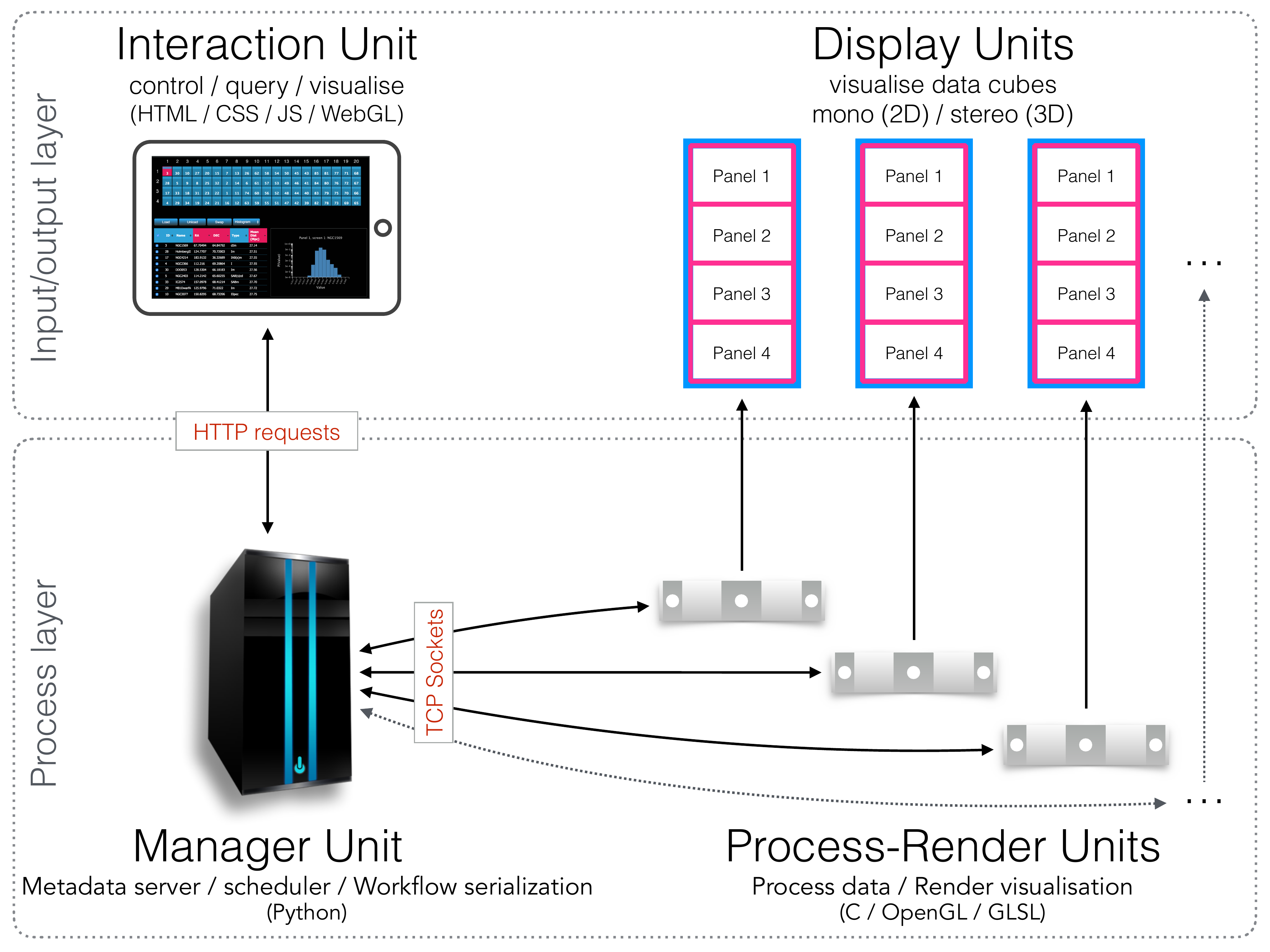}
\caption{Framework of {\tt encube} in the context of the Monash CAVE2. The {\em Interaction Unit} enables a user to control and query data displayed on the 80 stereo-capable screens of the Monash CAVE2. It can also render quantitative visualisations such as an histogram or a moment map on demand. The {\em Manager Unit} is in charge of serving data to the {\em Interaction Unit}, scheduling jobs, handling communications, and keeps a record of the workflow. The {\em Process-Render Units} process data (e.g. load spectral cubes, compute quantitative information), and render visualisations. Finally, the {\em Display Units} output spectral cubes visualisations for astronomers to see.}
\label{fig::framework}
\end{figure*}

{\tt encube} aims at enabling astronomers to interactively visualise, compare, and query a subset of spectral cubes from survey data. The framework includes several strategies for qualitative, quantitative, and comparative visualisation, including different mechanisms to organise, query and tag data interactively. {\tt encube} comprises two layers: the {\em Input/output layer}, and the {\em Process layer}. The {\em Input/output layer} includes an {\em Interaction unit} and multiple {\em Display Units}, while the {\em Process layer} includes a {\em Manager Unit} and multiple {\em Process-Render Units}. Figure \ref{fig::framework} depicts the framework and its related hardware and software components. For an in-depth description of {\tt encube} and its different units, see Vohl \etal\ (2016).

%is mapped to a touch-based controller (e.g. tablet, smart phone) or other portable devices
%The {\em Interaction unit} provides control over the global visualisation and analysis environment. It provides astronomers with different mechanisms to decide which spectral cubes to render, and where it should be rendered on the {\em Display Units} grid. It also enables to interactively modify how it one or many spectral cubes should be rendered (e.g. synchronised camera positions across all spectral cubes, transparency level, colormap). Other mechanisms include: querying the visualised data, and visualise quantitative information.

The {\em Manager Unit} is at the core of {\tt encube}. The {\em Manager Unit} has been designed to make it simple to incorporate other Python-based astronomical packages such as astropy (\cite{Astropy2013A&A...558A..33A}) and Virtual Observatory capabilities [e.g. TAP (\cite{2010ivoa.spec.0327D}), SAMP (\cite{Taylor201581})] developed within our community. In addition, as all communications and queries circulate through the \emph{Manager Unit}, it is where the workflow history record is being gathered.

Different interactive functionalities to compare spectral cubes are provided via the {\em Interaction Unit}. An {\em Interaction unit} provides control over the global visualisation and analysis environment. %It provides astronomers with different mechanisms to decide which spectral cubes to render, and where it should be rendered on the {\em Display Units} grid. 
Such functionalities include sorting spectral cubes, manually reordering displayed cubes, modifying rendering parameters, and querying one or many spectral cubes to obtain information like voxel distributions (i.e. histograms) or moment maps. Interaction with the many screens is provided through a web interface that can be used through different devices such as touch-based devices (tablet, smart phones) and other portable devices (e.g. laptops).  

The primary advantage of {\tt encube} comes from its concepts of `Single Instruction, Multiple Views' (SIMV) and `Single Instruction, Multiple Queries' (SIMQ). Sharing similarities to the concept of `Single Instruction, Multiple Data' (SIMD), these distributed models of processing and rendering allow one requested action to be applied to many data cubes in parallel. Hence, within the Monash CAVE2 environment for example, one can compare of order 100 spectral cubes.%, where the size of each cube is beyond that a typical workstation can handle. 
Through SIMV and SIMQ, instead of repeating a task (analysis or visualisation) over and over from one spectral cube to the next, one has the ability to spawn this task to multiple data cubes seamlessly.

% Section 3
%\section{Synchronous and asynchronous collaboration}
\section{Documenting the discovery workflow: a key for synchronous and asynchronous collaboration}
%In addition to mechanisms to interact with the visualisation (organise and query), {\tt encube} also contains strategies to serialize the workflow: actions taken throughout the discovery process are being recorded. [Hence, actions can be reviewed and reloaded.] either within the advanced visualisation environment or back at the researcher's desk. 

%Given the large number of displays available in the Monash CAVE2, users can execute many different tasks to many different files concurrently. 
To help researchers keep track of the discovery process, {\tt encube} includes different features to stay organised. In particular, {\tt encube} integrates the concept of workflow serialization: the generation of metadata about user interactions with the system. Two levels of interactions are considered for serialization: actions applied to a file, and actions applied on the overall distribution of visualisations. In addition, researchers can include annotations linked to specific data files by using searchable keywords called {\em tags}.

\subsection{Workflow serialization: data file level}\label{sec::worflow-file}
During a session, actions applied to data files can be stored for future evaluation. The type of information to be stored can be defined by the user. For example, one can keep track of visualisation parameters such as camera position (e.g. volume rotation angle, pan/zoom position), and volume rendering parameters (e.g. transparency, contrast). With data file level workflow, a user can review actions taken throughout the discovery process, either within the advanced visualisation environment or back at the researcher's desk. A user can restore a file to a given state in order to review interesting features, or simply continue the analysis process where a previous session ended. It also permits the discovery process to be replayed -- which may help the user to remember how and why certain steps have been taken.

\subsection{Workflow serialization: meta-visualisation level}\label{sec::worflow-meta}
Workflow at the meta-visualisation level keeps track of which data file is loaded at a given time, and where it is rendered (e.g. on which {\em Display Unit} it appears). This mechanism represents a way to store actions relative to the configuration for future evaluation. To provide a means for comparative visualisation in a collaborative fashion, a key feature of {\tt encube} is to rely on users' visual and spatial awareness. By keeping a record of the spatial configuration of visualisations within the visualisation environment (e.g. on which {\em Display Unit} a data file is rendered), it is possible for users to reproduce specific visualisation setups. 

\subsection{Workflow serialization: an example}\label{sec::worflow-example}
As a workflow example, consider a research team interested in the evaluation of galaxy kinematics from survey data. This team is formed of two sub groups, namely group A and B, where no members are shared between the two sub groups. Group A is available for a visit to the CAVE2 on Wednesday, while group B is available on Friday. On Wednesday, group A arranges galaxies into $N$ categories, and includes tags with each object, where each category shows similarities in kinematic morphology. On Friday, group B can evaluate all tags left behind by group A. By itself, a tag may be ambiguous. Hence, by reloading the overall configuration of visualisations within the CAVE2, group B can evaluate the tags in relation to the physical arrangement of galaxies on {\em Display Units}. This methodology provides a new asynchronous way of collaboration within teams.

\section{Current limitations \& on-going work}
Despite the considerable computational power available within the Monash CAVE2, there will be times where more computational resources will be required to visualise SKA-pathfinder scale spectral cubes. To cope with spectral cubes of terabyte (TB) scale (e.g. APERTIF and ASKAP cubes), the distributed visualisation and analysis framework {\tt GraphTIVA} was introduced (\cite{Hassan2013}).  For a 0.5 TB spectral cube using a cluster of 96 GPUs, \cite{Hassan2013} showed volume rendering at 7--10 frames per second could be achieved; and computation of basic global image statistics such as the mean intensity and standard deviation in 1.7 seconds. 

%We are currently proceeding with the evaluation of the integration of {\tt GraphTIVA} into AWS cloud computing\footnote{https://aws.amazon.com/} through the SKA/AWS AstroCompute in the Cloud grant scheme (Hassan \etal\ in prep.). A cloud computing solution could, in principle, provide scalability to a user in order to have access to any number of GPUs, while requiring minimal installation and setup by the end user. In addition, w
We are evaluating the possibility of integrating {\tt GraphTIVA} within the {\tt encube} framework to provide HI astronomers with novel way to interactively analyse TB scale spectral cubes' sources within their environment, at high resolution via advanced visualisation environments like the Monash CAVE2. This would offer a mode of operation where a full resolution spectral cube is rendered on a $4 \times 4$ grid of displays, leaving $16 \times 4=64$ screens for individual sources found within the parent cube. 

\section{Final thoughts}
As we are entering the Petascale Astronomy Era and its large scale spectral cube surveys, there is a need for novel methods to be explored. Our visual analytics framework has the potential to empower researchers with ways to quickly manipulate and visualise large subsets of their data. To this end, the CAVE2 permits new approaches to, and applications of, visual analytics. It offers great potential to accelerate the discovery process in the era of large-scale spectral-cube surveys.  {\tt encube} provides a best of both worlds approach through support of high-end, collaborative visualisation in the CAVE2 while also supporting desktop-based analysis and discovery. We note that {\tt encube} is not only viable for astronomy, but for any volumetric scientific data surveys (e.g. medical imaging, earth sciences). 

\acknowledgements 
This work was enabled and supported by the Monash Immersive Visualisation Platform (http://monash.edu/mivp). DV acknowledges the support of the Astronomical Society of Australia for travel funding.

\end{document}